\documentclass[aps, pre, preprint,unsortedaddress,a4paper,onecolumn, showkeys]{revtex4}

\usepackage[]{amsmath}
\usepackage{amssymb}
\usepackage[dvips]{graphicx}
\usepackage{color}
\usepackage{tabularx}
\usepackage{mathtools}
\usepackage{algpseudocode}
\usepackage{enumitem}
\usepackage{multirow}
\usepackage{epstopdf}
\usepackage{bm}
\usepackage{tikz}
\usepackage{subfig}
\usepackage{upgreek}
\usetikzlibrary{shapes,snakes}
\usetikzlibrary{arrows}
\usepackage[normalem]{ulem}
\usepackage{hyperref}

 \newcommand{\MC}[1]{{\color{black} #1}}

\begin{document}

\title{
\MC{DP Compress: a Model Compression Scheme for Generating Efficient Deep Potential Models}
  }

\author{Denghui Lu}
\affiliation{HEDPS, CAPT, College of Engineering, Peking University, Beijing 100871, P.R.~China}

\author{Wanrun Jiang}
\affiliation{Songshan Lake Materials Laboratory, Dongguan, Guangdong 523808, P.R.~China}
\affiliation{Institute of Physics, Chinese Academy of Sciences, Beijing 100190, P.R.~China}

\author{Yixiao Chen}
\affiliation{Program in Applied and Computational Mathematics, Princeton University, Princeton, NJ, USA}

\author{Linfeng Zhang}
\affiliation{Beijing Institute of Big Data Research, Beijing 100871, P.R.~China}

\author{Weile Jia}
\affiliation{Institute of Computing Technology, Chinese Academy of Sciences, Beijing 100190, P.R.~China}
\affiliation{University of Chinese Academy of Sciences, Beijing 100049, P.R.~China}

\author{Han Wang}
\affiliation{Laboratory of Computational Physics,
  Institute of Applied Physics and Computational Mathematics, Fenghao East Road 2, Beijing 100094, P.R.~China}
\affiliation{HEDPS, CAPT, College of Engineering, Peking University, Beijing 100871, P.R.~China}

\author{Mohan Chen}
\email{mohanchen@pku.edu.cn}
\affiliation{HEDPS, CAPT, College of Engineering, Peking University, Beijing 100871, P.R.~China}

\begin{abstract}
Machine-learning-based interatomic potential energy surface (PES) models are revolutionizing the field of molecular modeling.
However, although much faster than electronic structure schemes,
\MC{these models suffer from costly computations via deep neural networks to predict the energy and atomic forces,
resulting in lower running efficiency as compared to the typical empirical force fields.}
Herein, we report a model compression scheme for boosting the performance of the Deep Potential (DP) model, a deep learning based PES model.
This scheme, we call DP Compress, is an efficient post-processing step after the training of DP models (DP Train).
DP Compress combines several DP-specific compression techniques, which typically speed up DP-based molecular dynamics simulations by an order of magnitude faster, and consume an order of magnitude less memory.
We demonstrate that DP Compress is sufficiently accurate by testing a variety of physical properties of Cu, H$_2$O, and Al-Cu-Mg systems.
DP Compress applies to both CPU and GPU machines and is publicly available \MC{online}.
\end{abstract}

\MC{\keywords{Deep Potential Molecular Dynamics, Model Compression, Machine Learning}}

\maketitle

\setlength{\parskip}{.5em}

\section{Introduction}
Deep learning is leading to a paradigm shift in molecular dynamics (MD), the {\it de facto} lens for the microscopic understanding of a broad spectrum of issues, such as drug discovery, complex chemical processes, nanotechnology, etc.
Some protocols that integrate physics-based principles and advantages of deep neural networks (DNN), while retaining the accuracy of quantum mechanics (or {\it ab initio}) models, can greatly boost the accessible time and size scales by several orders of magnitude~\cite{behler2007generalized,chmiela2017machine,schutt2017schnet,smith2017ani,zhang2018deep,zhang2018end,zhang2020efficient,zhang2021accelerating}.
Recently, a highly optimized implementation of linear-scaling Deep Potential Molecular Dynamics (DeePMD), a deep learning-based MD scheme, has pushed the limit of molecular dynamics with {\it ab initio} accuracy to 100 million atoms, with a computational cost that typically requires one day for nano-second (ns) simulations~\cite{LU2020CPC,Jia2020GB}.
DeePMD has enabled various applications in, for example,
reactive uptake of nitrogen oxides by aqueous aerosol~\cite{galib2021reactive},
crystal nucleation of liquid silicon~\cite{bonati2018silicon},
liquid-liquid phase transition of water~\cite{gartner2020signatures},
one dimensional cooperative diffusion in three dimensional crystal~\cite{wang2021electronically},
structural order in quasicrystal growth~\cite{han2020dynamic},
phase diagram of water~\cite{zhang2021phase},
and warm dense matter~\cite{zhang2020pop,liu2020jpcm},
etc.
However, for many important problems that require large system sizes or long time scales, the estimated computational cost is still prohibitive.
For example, in order to perform a 1-ns simulation of 1 million Cu atoms using the DeePMD-kit code~\cite{Jia2020GB} on a machine with 150 V100 GPU cards, about half a week would be needed.
Therefore, to truly make large-scale molecular simulation with {\it ab initio} accuracy a routine procedure, a major demand is to improve the efficiency of model inference, while reducing the computational cost.

In conventional deep learning territories like speech recognition, visual object recognition, object detection, etc., it has become a common practice to conduct model compression\MC{~\cite{elsken2019neural,choudhary2020comprehensive,xia2020efficient}} between two stages: model training and model inference.
\MC{The model architecture has to be carefully designed, typically in the form of large and deep neural networks, so that the parameters are easy to optimize.~\cite{li2012brief, buscema1998back}}
\MC{On the one hand, the model architecture has to be carefully designed, typically in the form of large and deep neural networks, so that the parameters are easy to optimize.~\cite{li2012brief, buscema1998back} Regarding the DeePMD method, the designing principles of the model architecture have been described in previous works~\cite{zhang2018deep,wang2018kit}. On the other hand, after training the model,}
the computational cost for model inference may be greatly reduced without significant drop in accuracy
\MC{by using model compression methods such as parameter pruning, low-rank factorization, weight quantization, as well as some judiciously designed neural architecture search and grow-and-prune schemes~\cite{sainath2013low,han2015deep,hubara2017quantized,zoph2016neural,li2020train}.}
We refer to recent review articles~\cite{elsken2019neural,choudhary2020comprehensive,xia2020efficient} for thorough discussions of \MC{the related issues in the model compression methods}.

Two issues need to be addressed when the model compression techniques are used for deep learning assisted physical models.
We take Deep Potential (DP)~\cite{zhang2018deep,zhang2018end}, the interatomic potential energy model for driving DeePMD simulations, as an example.
First, the accuracy requirement for DP is much more strict than conventional model compression tasks.
For a system, the energy and the forces, and hence many calculated properties, predicted by the compressed DP model should show negligible differences with the original model.
Second, the network structure is not as deep and large as in conventional deep learning tasks.
Therefore, many existing compression techniques are not directly applicable.

In this work, we introduce DP Compress, an efficient post-processing step after the training of DP models (DP Train), the resulting compressed model can be directly utilized in atomistic simulations.
DP Compress combines several DP-specific compression techniques, which typically speedup the state-of-the-art DeePMD code by an order of magnitude faster, and an \MC{order of magnitude more memory-efficient}.
We demonstrate that the error induced by the compression process is sufficiently small, and such an improvement applies to both CPU and GPU machines, so that users with different hardware environments can benefit from it.
In practice, we generate compressed models for Cu, H${}_2$O and Al-Cu-Mg alloy systems, and investigate their performances on the predictions of various physical properties, using both CPU and GPU machines.
Moreover, to provide the users with a better guidance, we study in detail how the performance changes with the architectures of the neural network model.

The DP Compress scheme has been implemented in the open-source package DeePMD-kit~\cite{wang2018kit},
which is written in Python and C++.
For efficient and flexible training, the code has been interfaced with major machine learning frameworks like TensorFlow~\cite{abadi2016tensorflow}.
For large-scale MD simulations, it has been interfaced with popular MD software like LAMMPS~\cite{thompson2022lammps}.
Using DP Compress, with a simple command, the users of DeePMD-kit can typically gain an order of magnitude efficiency when performing DeePMD simulations.

\section{Methods}


\begin{figure}
  \centering
  \includegraphics[width=0.96\textwidth]{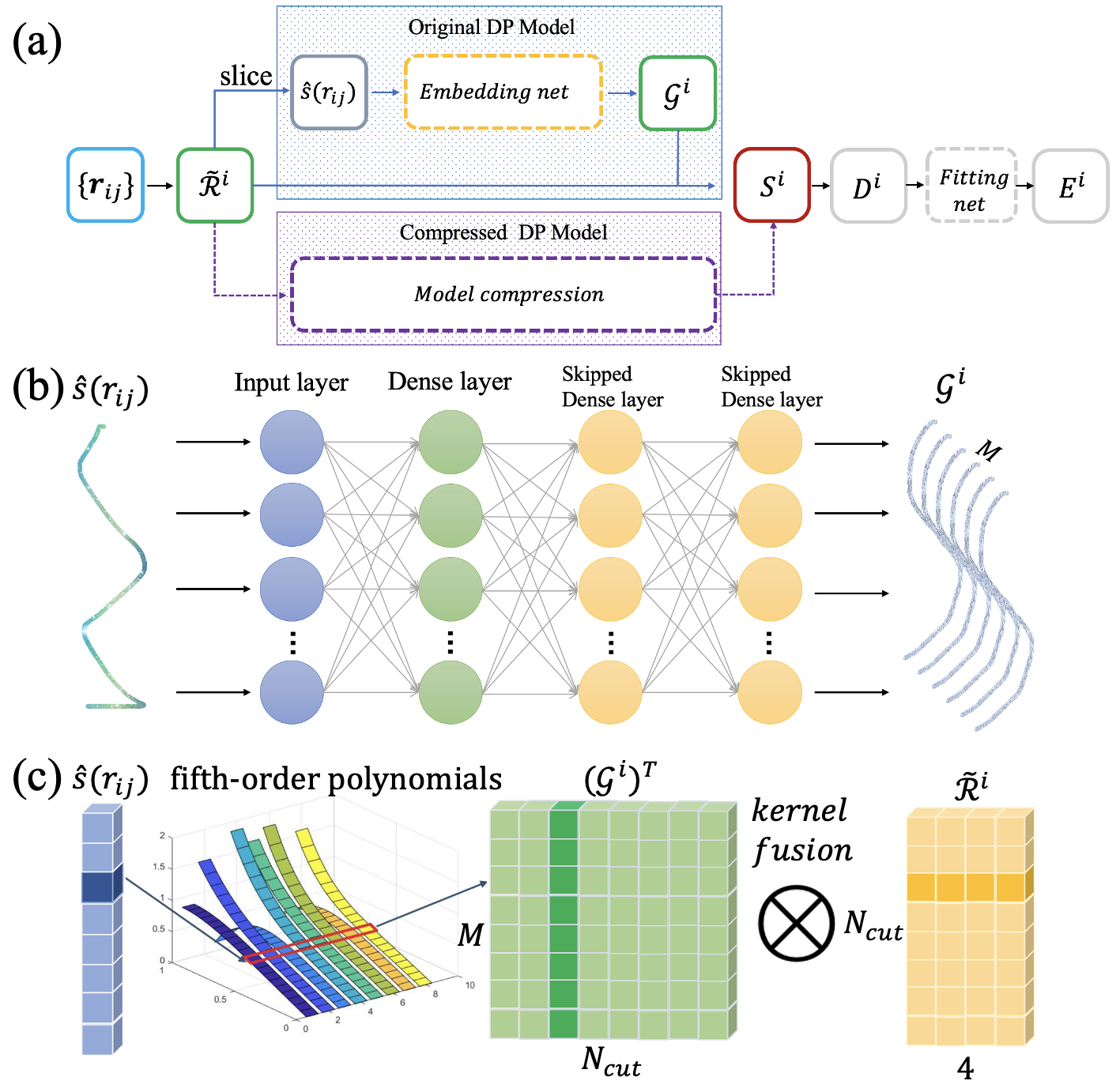}
  \caption{\label{fig:workflow}
(a) Schematic plot of the single-atom workflow in the original and compressed DP models. $\{\bm{r}_{ij}\}$ depicts the coordinates of neighboring atoms $j$ relative to atom $i$. The environment matrix of atom $i$ is denoted as $\tilde{\mathcal R}^i$, which has dimensions of $N_{cut}\times4$ with $N_{cut}$ being the cutoff number.
(b) In the original DP model, the {\it embedding net} maps the normalized weighting function $\hat{s}(r_{ij})$\MC{, sliced from the first column of $\tilde{\mathcal R}^i$,} to the local embedding matrix $\mathcal G^i$, which leads to the ${\mathcal S}^i=\mathcal (\mathcal G^{i})^{T}{\tilde{\mathcal R}}^i$ matrix that has dimensions of $M\times4$. The resulting descriptor $D_i$ is mapped by the {\it fitting net} to the energy $E^i$.
(c) In the compressed DP model, the {\it embedding net} operations are \MC{replaced by calculations based on tabulated fifth-order polynomials}, enabling efficient computations of ${\mathcal S}^i$ via the kernel fusion operation.
}
\end{figure}

We divide the typical DeePMD workflow into four steps.
First, the preparation of the training data, which consists of a set of atomic types and atomic coordinates, as well as corresponding labels (energies, forces, \MC{and optionally, virials}) obtained from quantum mechanical calculations.
Note that an advanced scheme to generate an optimal set of training data involves a concurrent learning procedure that \MC{iteratively conducts} model training, DeePMD exploration, and quantum mechanics calculations (See, e.g.,  Refs.~\cite{zhang2019active,zhang2020dpgen} for details).
\MC{The second step is the model training procedure with the training data, which optimizes the neural network parameters in a DP model.}
The typical training time spans from several hours to one week on a single GPU card, depending on the complexity of the data.
Third, the model freezing process identifies and saves all of the required model information, including the computational graph, neural network parameters, etc., in a single file.
Finally, the frozen model can be used for DeePMD simulations, during which the interatomic energies, forces, and/or virials, are calculated on-the-fly using the frozen model.

\MC{The total interatomic energy $E$ is constructed as the summation of contributions from all of the atomic energy contributions $E^i$, i.e., $E=\sum_i E^i$.
The procedure of computing the the atomic energy $E_i$ from the relative coordinates between atom $i$ and its near neighbors is referred as ``single-atom workflow"\MC{\cite{behler2007generalized}}.}
We use $\mathcal{N}_i=\{j|r_{ij}<r_c\}$ to denote the set of neighboring atoms of atom $i$ within a real-space cutoff $r_c$, and $r_{ij}$ denotes the distance between atoms $i$ and $j$.
Importantly, the DP model preserves necessary symmetry properties of a PES model, i.e., translational symmetry, rotational symmetry, as well as permutational symmetry, and it is fully end-to-end\MC{~\cite{zhang2018end}}.
%
Note that for simplicity, here we only introduce the procedure to compute $E^i$, but similar considerations apply to the calculations of the forces and the virials, which involve the derivatives of $E^i$ with respect to $i$ and $j$.

We first construct an environment matrix $\tilde{\mathcal{R}}^i$ to characterize the environment of a given atom $i$,
which is illustrated in Fig.~\ref{fig:workflow}(a).
\MC{In order to obtain $\tilde{\mathcal{R}}^i$, for each pair distance $r_{ij}$,  we introduce a weighting function $s(r_{ij})$ with the form of}
\MC{\begin{align}
    s (r_{ij}) = \left \{
    \begin{aligned}
    &\frac{1}{r_{ij}} && r_{ij} < r_{cs} \\
    &\frac{1}{r_{ij}}\,[ u^3 (-6u^2 + 15 u - 10 ) + 1] && r_{cs} \leq r_{ij} < r_c\\
    &0 && r_c \leq r_{ij}
    \end{aligned}
    \right., \quad u = \frac{r_{ij}-r_{cs}}{r_c - r_{cs}}. \label{equ21} 
\end{align}}
\MC{Here $r_{cs}$ is a smooth cutoff that allows $\tilde{\mathcal{R}}^i$ to smoothly decay to zero} as $r_{ij}$ approaches the cutoff $r_c$.
The use of $s(r_{ij})$ enforces a continuous evolution when atoms enter/exit the neighborhood of $i$ by describing the atomic coordinates $\bm{r}_{ij}=(x_{ij}, y_{ij}, z_{ij})$ with a 4-vector $(s(r_{ij}), s(r_{ij})x_{ij}, s(r_{ij})y_{ij}, s(r_{ij})z_{ij})$,
which then undergoes a normalization procedure.
\MC{The normalization procedure is explained as follows: First,
the data sampled from the training set. Second, we calculate the mean value and standard deviation of each element in the 4-vector. To be specific,
for the first element $s(r_{ij})$, we subtract the mean value $\overline{s(r_{ij})}$ and divide the result by the standard deviation of $s(r_{ij})$ , which is labelled as $\sigma(s(r_{ij}))$. Third, we collect all of the data and denote the whole map as $\hat{s}(r_{ij}):$\begin{equation}
  \hat{s}(r_{ij}) = \frac{s(r_{ij}) - \overline{s(r_{ij})}}{\sigma (s(r_{ij})).}
\end{equation}
For the remaining three elements, since they are equivalent under arbitrary spatial rotation operations, we divide them by their overall standard deviation $\sigma_{all}$. Note that $s(r_{ij})x_{ij},\ s(r_{ij})y_{ij}$ and $s(r_{ij})z_{ij}$ are put together to calculate the standard deviation $\sigma_{all}$:
\begin{equation}
  \sigma_{all} = \sigma(\{{s(r_{ij})x_{ij}},  {s(r_{ij})y_{ij}}, {s(r_{ij})z_{ij}}\}).
\end{equation}
After the above procedures, we denote the resulting 4-vector as $\tilde{\bm{r}}_{ij}$ with the form of
\begin{equation}
  \tilde{\bm{r}}_{ij} = \left(\hat{s}(r_{ij}), 
  \frac{s(r_{ij})x_{ij}}{\sigma_{all}},
  \frac{s(r_{ij})y_{ij}}{\sigma_{all}},
  \frac{s(r_{ij})z_{ij}}{\sigma_{all}}
  \right),
\end{equation} 
which is then used to define the environment matrix $\tilde{\mathcal{R}}^i$.}

Next, a descriptor $\mathcal D^i$ is generated based on the environment matrix $\tilde{\mathcal{R}}^i$. First,
in the original single-atom workflow of DeePMD-kit, the rows of $\tilde{\mathcal R}^i$ are mostly given by $\tilde{\bm{r}}_{ij}$, with the neighbors $j$ sorted
first according to their types $\alpha$, and then according to their distances to $i$.
Meanwhile, one needs to set the maximal number of neighbors of each atom type, denoted as $sel_\alpha$. As a result, the number of rows of $\tilde{\mathcal R}^i$, we call the cutoff number, is fixed to be $N_{cut}=\sum_\alpha sel_\alpha$.
If the number of neighbors of type $\alpha$ is smaller than $sel_\alpha$, the remaining rows of $\tilde{\mathcal R}^i$ for that type are fed with zeros.
Second, the embedding matrix ${\mathcal G}^i$ = $(G^i_{jm})$ = $( G^i_m (\hat{s}(r_{ij}) ))$, with $N_{cut}$ rows and $M$ columns, is generated by a DNN named \emph{embedding net}.
Third, multiplication of $({\mathcal G}^i)^T$ by ${\tilde{\mathcal R}}^i$ yields the matrix ${\mathcal S}^i=({\mathcal G}^i)^T{\tilde{\mathcal R}}^i$ with $M$ rows and 4 columns.
Let ${\mathcal S}^{i<}$ be the matrix formed by the first $M'$ ($<M$) rows of ${\mathcal S}^i$.
\MC{In practice, we found a small $M'$ value still preserves the accuracy and reduces the computational costs.}
Finally, the descriptor is generated via $\mathcal D^i={\mathcal S}^i({\mathcal S}^{i<})^T$ with $M$ rows and $M'$ columns.
The descriptor is then passed to a {\it fitting net}, a fully connected DNN, which outputs the atomic energy contribution $E^i$.

Now, we describe the process to compress the single-atom workflow after training.
First of all, \MC{during the model inference procedure, the calculations of the embedding and $\mathcal S^i$ matrices are the most computationally expensive parts} in both calculations of the forward and the backward propagation.  
Typically, more than 80\% of the computational time  \MC{(detailed in section III)} is spent on this part by using either CPU or GPU machines. 
Essentially, by using the $\hat{s}(r_{ij})$ as inputs, the trained embedding network learns a list of 1-dimensional functions.
Therefore, the embedding network can be substituted with tabulated functions to largely speedup the computations.
Here we adopt the \MC{Hermite interpolation method and use} piecewise fifth-order polynomials to interpolate the embedding functions.
Specifically, the multi-valued embedding network can be replaced by a set of multiple single-valued functions: ${\mathcal G_{m}}$, $m=1,2,\dots,M$.
We divide the domain of the input tensor $\hat{s}(r_{ij})$ into $L$ equally spaced components, and the $L+1$ interpolation points are labelled as $x_1, x_2, ..., x_l, ..., x_{L+1}$.
\MC{The value of ``$L$" can be chosen to yield a small value of tabulation step $\Delta t$ that meets the accuracy requirements.}
Within the region [$x_{l}, x_{l+1}$), we define a fifth-order polynomial \MC{$g_{m}^{l}(x)$} according to the following formula:
\begin{equation}
    \MC{g_{m}^{l}(x)} = a_{m}^{l}x^{5} + b_{m}^{l}x^{4} + c_{m}^{l}x^{3} + d_{m}^{l}x^{2} + e_{m}^{l}x + f_{m}^{l},
\end{equation}
where $a_{m}^{l}$, $b_{m}^{l}$, $c_{m}^{l}$, $d_{m}^{l}$, $e_{m}^{l}$, and $f_{m}^{l}$ are fitting parameters.

In addition, we prepare six constraints at the two mesh points $x_{l}$ and $x_{l+1}$ in order to compute the above six coefficients. To be specific, for each mesh point, we compute the value of the embedding function
\begin{equation}
    y_l = {\mathcal G_{m}}(x_{l}),\label{con:value}
\end{equation}
the first-order derivative
\begin{equation}
y^{'}_l = {\mathcal G^{'}_{m}}(x_{l}),\label{con:first-derivative}
\end{equation}
and the second-order derivative
\begin{equation}
y^{''}_{l} = {\mathcal G^{''}_{m}}(x_{l}).\label{con:second-derivative}
\end{equation}

The resulting formulas for the six coefficients can be written as
\begin{equation}
a_{m}^{l}=\frac{1}{2\MC{\Delta}t^5}[12h-6(y^{'}_{l+1}+y^{'}_{l})\MC{\Delta}t+(y^{''}_{l+1}-y^{''}_{l})\MC{\Delta}t^2],
\end{equation}
\begin{equation}
b_{m}^{l}=\frac{1}{2\MC{\Delta}t^4}[-30h+(14y^{'}_{l+1}+16y^{'}_{l})\MC{\Delta}t+(3y^{''}_{l}-2y^{''}_{l+1})\MC{\Delta}t^2],
\end{equation}
\begin{equation}
c_{m}^{l}=\frac{1}{2\MC{\Delta}t^3}[20h-(8y^{'}_{l+1}+12y^{'}_{l})\MC{\Delta}t-(3y^{''}_{l}-y^{''}_{l+1})\MC{\Delta}t^2],
\end{equation}
\begin{equation}
d_{m}^{l}=\frac{1}{2}y^{''}_{l},
\end{equation}
\begin{equation}
e_{m}^{l}=y^{'}_{l},
\end{equation}
\begin{equation}
f_{m}^{l}=y_{l}.
\end{equation}
where \MC{the tabulation step} $\MC{\Delta}t=x_{l+1}-x_l$ and $h=y_{l+1} - y_l$. In practice, both $\mathcal{G}^{'}_{m}(x_{l})$ and $\mathcal{G}^{''}_{m}(x_{l})$ can be efficiently evaluated, which enables the model compression process to be finished within a few minutes on a single-node CPU machine.

The accuracy of the compressed DP model depends on the tabulation step $\Delta t$, while the range of $r_{ij}$ being interpolated is determined by the training data. For example, the lower bound of $r_{ij}$, which is denoted as $r_l$, is chosen by scanning all of the training data.
In addition, the algorithm guarantees that the upper bound of $r_{ij}$ used in the compression model is larger than the maximum value found in the training data.
In this regard, the range of $\hat{s}(r_{ij})$ being interpolated is [$\hat{s}(r_c)$, $\hat{s}(r_l)$].
As shown in Fig.~\ref{fig:performance}(a), with a tabulation step of 10${}^{-2}$, the errors of energy and force can be controlled below 10${}^{-3}$ meV/atom and 10${}^{-4}$ meV/\AA, respectively, much lower than the typical training errors of the DP model.
Therefore, we set the default tabulation step to be 10${}^{-2}$.

In the original DP workflow,
the embedding matrices from all of the atoms typically consume more than $90$ percent of the total host/device memory usage~\cite{LU2020CPC}, which becomes a bottleneck for simulating a larger number of atoms.
Specifically, in the single-atom workflow, the embedding matrices are loaded to the registers and the matrix product ${\mathcal S}^i=({\mathcal G}^i)^T{\tilde{\mathcal R}}^i$ is computed.
The above process causes a huge data movement overhead between the registers and the host/device memory, which is memory-bound by the host/device memory throughput\cite{Jia2020GB}.
In the compressed DP workflow illustrated in Fig.~\ref{fig:workflow}(c),
after tabulating the embedding matrix ${\mathcal G}^i$ with fifth-order polynomials, we perform an efficient kernel fusion step to yield ${\mathcal S}^i$.
Taking the GPU implementation as an example, for each atom $i$, the matrix multiplication ${\mathcal S}^i=({\mathcal G}^i)^T{\tilde{\mathcal R}}^i$ is handled by a single thread block. In detail, when one column of $(\mathcal{G}^i)^T$ is evaluated and stored in registers (without storing back to global memory),
 the corresponding row of the environment matrix $\tilde{\mathcal{R}^i}$ is loaded into the register to perform an outer-product with the column of $(\mathcal{G}^i)^T$. The outer-product has to be performed at most $N_{cut}$ times to yield ${\mathcal S}^i$, which has dimensions of $M\times 4$. 
We remark that $\mathcal{G}^i$ neither allocated nor moved between global memory and registers in  the optimized code, and both the memory footprint and computational time are significantly reduced after the kernel fusion.

In a practical simulation, the number of neighbors with type $\alpha$ can be much smaller than $sel_\alpha$,  so that the environment matrix $\tilde{\mathcal R}^i$ may have a large number of redundant zeros.
This issue is particularly serious when one trains a DP model with data in a large concentration range for a few types of atoms, so that $sel_\alpha$ has to be very large for each type.
{In the fused kernel of matrix product ${\mathcal S}^i=({\mathcal G}^i)^T{\tilde{\mathcal R}}^i$,
the column of $({\mathcal G}^i)^T$ is evaluated and the following outer product is performed only when $j$ is a valid neighbor of atom $i$.
This fine-grained and conditional matrix production is not possible in the original DP workflow, in which the matrix product is performed by a GEMM \MC{(General Matrix Multiplication)} call.}
{As a result, these operations reduce the floating operations per second (FLOPS) and data access at the same time.}


\section{Results and Discussion}
We adopt three well-benchmarked systems, i.e., Cu, H${}_2$O, and Al-Cu-Mg, to validate the DP Compress scheme. First, for the Cu system, the authors in Ref.~\cite{zhang2020dpgen} adopted a concurrent learning scheme~\cite{zhang2019active} to prepare an optimal set of {\it ab initio} training data and generated a Cu model with an uniform accuracy over a wide range of thermodynamic conditions (temperatures up to $\sim$2600 K and pressures up to $\sim$5 GPa).
Second, for the H${}_2$O system, previous works in Refs.~\cite{zhang2018deep,ko2019isotope} have shown that DeePMD can accurately capture the delicate balance between weak non-covalent intermolecular interactions, thermal effects, as well as nuclear quantum effects in water.
Extensions of the DP formulation have made possible accurate predictions of the infrared~\cite{zhang2020dw} and Raman~\cite{grace2020raman} spectra of water.
More recently, part of the authors have generated a DP model to study the phase diagram of water ranging from low temperatures to about 2400 K and low pressures to 50 GPa, excluding the vapor stability region~\cite{zhang2021phase}.
Last, for the Al-Cu-Mg system in the full concentration range, a DP model was generated that yields predictions consistent with first-principles calculations for various binary and ternary systems on their fundamental energetic and mechanical properties~\cite{jiang2021accurate}.

We list the parameters utilized in the calculations. \MC{For Cu~\cite{zhang2020dpgen,Jia2020GB}, H${}_2$O~\cite{zhang2021phase}, and Al-Cu-Mg~\cite{jiang2021accurate}}, the cutoff radii $r_c$ (number of neighbors $N_{cut}$) are chosen to be 8.0~\AA~(512), 6.0~\AA~(144), and 9.0~\AA~(1800), respectively.
\MC{The cutoff radii are the same as those reported in the corresponding original publications.}
The size of the {\it fitting nets} are set to ($240, 240,240$), while the sizes of the {\it embedding nets} are ($32,64,128$) for Cu and H${}_2$O systems and ($25,50,100$) for Al-Cu-Mg systems.
More details of these models and training data can be found in \MC{the Supporting Information}.
While using the above protocol for the following tests, we also study the influences of different setups on the final results, including the type of precision, the length of the tabulation step, as well as the network structure.
All of the calculations are performed with the LAMMPS package~\cite{thompson2022lammps} that has an interface with DeePMD-kit.

\begin{figure}
  \centering
  \includegraphics[width=0.96\textwidth]{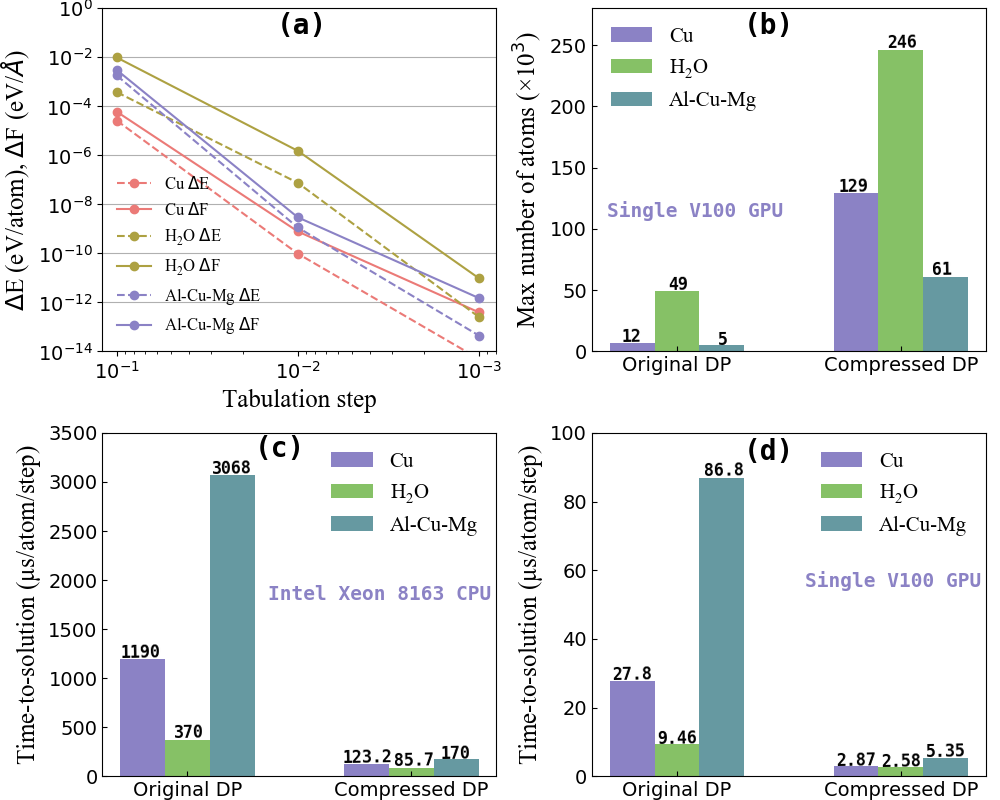}
  \caption{\label{fig:performance}
Performances of the original and compressed DP models, tested on three systems: Cu, H${}_2$O, and Al-Cu-Mg.
  (a) Root mean square errors of energy and forces as functions of the tabulation step;
  (b) Maximal number of atoms that can be simulated on an NVIDIA V100 GPU (32 GB memory);
  (c) Time-to-solution ($\upmu$s/atom/step) of DP models tested on a 6-core Intel Xeon 8163 CPU;
  (d) Time-to-solution ($\upmu$s/atom/step) of DP models tested on an NVIDIA V100 GPU.
  The Cu, H${}_2$O, and Al-Cu-Mg systems being tested for (a), (c), and (d) contain 6912,
  12288, and 5120 atoms, respectively.
  For producing (c) and (d), the MD equations are numerically integrated for 500 steps (the energy and forces are evaluated for 501 times).
  }
\end{figure}

Fig.~\ref{fig:performance} illustrates the performances of the original and compressed models on the Cu, H$_2$O, and Al-Cu-Mg systems.
\MC{By using the original DP model, we find 94.2\%, 87.5\% and 97.3\% of the forward and the backward propagation time is spent on the calculations of the $\mathcal G^i$ and $\mathcal S^i$ matrices in the Cu, H$_2$O, and Al-Cu-Mg systems, respectively.}
In Fig.~\ref{fig:performance}(a), we adopt three tabulation steps, i.e., $\Delta t=$0.1, 0.01, and 0.001, to generate three compressed models for each system.
We observe that the energy deviation $\Delta E$ and force deviation $\Delta F$ decrease as $\Delta t$ becomes smaller, and conclude that $\Delta t=$0.01 is accurate enough for most DeePMD applications.
In Fig.~\ref{fig:performance}(b), we compare the number of atoms that can be simulated on one single V100 GPU by using the original and compressed models. By using the compressed model, we find that the number of atoms in the Cu/H${}_2$O/Al-Cu-Mg system increases by roughly an order of magnitude. For example, the maximal number of atoms \MC{calculated} by one GPU increases from 12/49/5 thousands to 129/246/61 thousands for the Cu/H${}_2$O/Al-Cu-Mg system, respectively.
Because the compressed DP model does not store the local embedding matrix, DeePMD is able to handle a larger number of atoms on a given machine.
Meanwhile, the compressed model \MC{significantly} speeds up the single-atom workflow. Figs.~\ref{fig:performance}(c) and (d) show that the DeePMD simulations with the compressed models are \MC{much} faster than the original models on CPU and GPU machines, respectively.
For instance, the speedups for Cu, H$_2$O, and Al-Cu-Mg are 9.69, 3.67, and 16.22 on a single V100 GPU, respectively.

\begin{figure}
 \centering
 \includegraphics[width=0.96\textwidth]{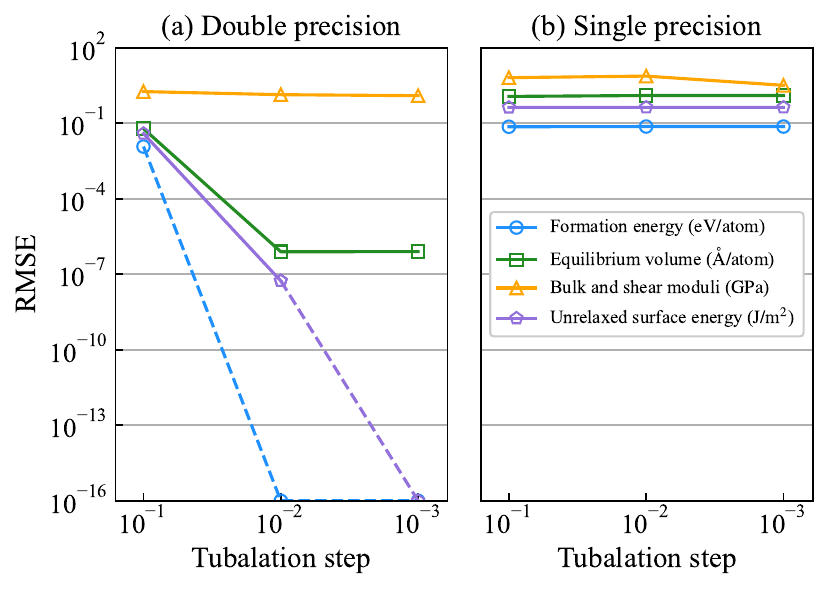}
 \caption{\label{fig:rmse-properties}
Root-mean-squared error (RMSE) of four properties for 6 Cu structures and 58 binary and ternary Al-Cu-Mg alloys. The properties include the formation energy, the equilibrium volume, the bulk and shear moduli, and the unrelaxed surface energy. Three tabulation steps are used with both double and single precisions. Note that we set the lower bound of errors to be $10^{-16}$
by considering the limit of significant digits under double precision, and the resulting points are connected with dashed lines.
}
\end{figure}

We summarize two major factors that affect the performance of DP Compress and its improvement over the original model.
First, the \MC{maximum number ($N_{cut}$)} of neighbors (H${}_2$O$<$Cu$<$Al-Cu-Mg) depends on the cutoff radius $r_c$ and the density of a system;
with similar network architectures, this factor roughly determines the maximal number of atoms that can be simulated on a given machine (H${}_2$O$>$Cu$>$Al-Cu-Mg in Fig.~\ref{fig:performance} (b)), as well as the ultimate time-to-solution of the compressed model (H${}_2$O$<$Cu$<$Al-Cu-Mg in Figs.~\ref{fig:performance} (c-d)).
Second, the difference between the cutoff number $N_{cut}$ and the average number of neighbors also plays an important role in determining the speed-up ratio.
\MC{In the case of water, the $N_{cut}$ and the average number of neighbors are 144 and 88, respectively, while those of the Al-Cu-Mg system are 1800 and 171, respectively.
This explains} why on both CPU and GPU, the speed-up ratio for H${}_2$O is much smaller than that for Al-Cu-Mg. 
On one hand, the H${}_2$O model is trained from snapshots of liquid water at ambient conditions, so the local density fluctuation is small in water and $N_{cut}$ is close to the average number of neighbors.
On the other hand, the trained DP model for Al-Cu-Mg system covers the full concentration range, 
\MC{so $sel_\alpha$ has to be large enough (600) to cover the highest density cases of pure metal configurations.}
Therefore, in practical simulations, the average number of neighbors is much smaller than $N_{cut}$ and a substantial amount of redundant zeros exist in the environment matrix $\tilde{\mathcal{R}}^i$; this redundant problem is well addressed by the kernel fusion procedure in the compressed model.

To validate the accuracy and applicability of the DP Compress scheme, we first evaluate the compressed model by computing several physical properties of the Cu and Al-Cu-Mg systems and compare them to the original model.
Specifically, we study four energetic and mechanical properties of the \MC{Cu and }Al-Cu-Mg systems: the formation energies, the equilibrium volumes, the elastic moduli, and the unrelaxed surface formation energies.
In these test, we collect 6 crystal structures of Cu and 58 crystal structures of binary and ternary alloys consisting of Al, Mg and Cu elements; these structures are taken from the Materials Projects (MP) database~\cite{jain2013commentary}. Note that the test set covers all of the documented crystals under corresponding elements combinations, and most of them are not explicitly covered by the training sets of the DP model.

\begin{figure}
 \centering
 \includegraphics[width=0.96\textwidth]{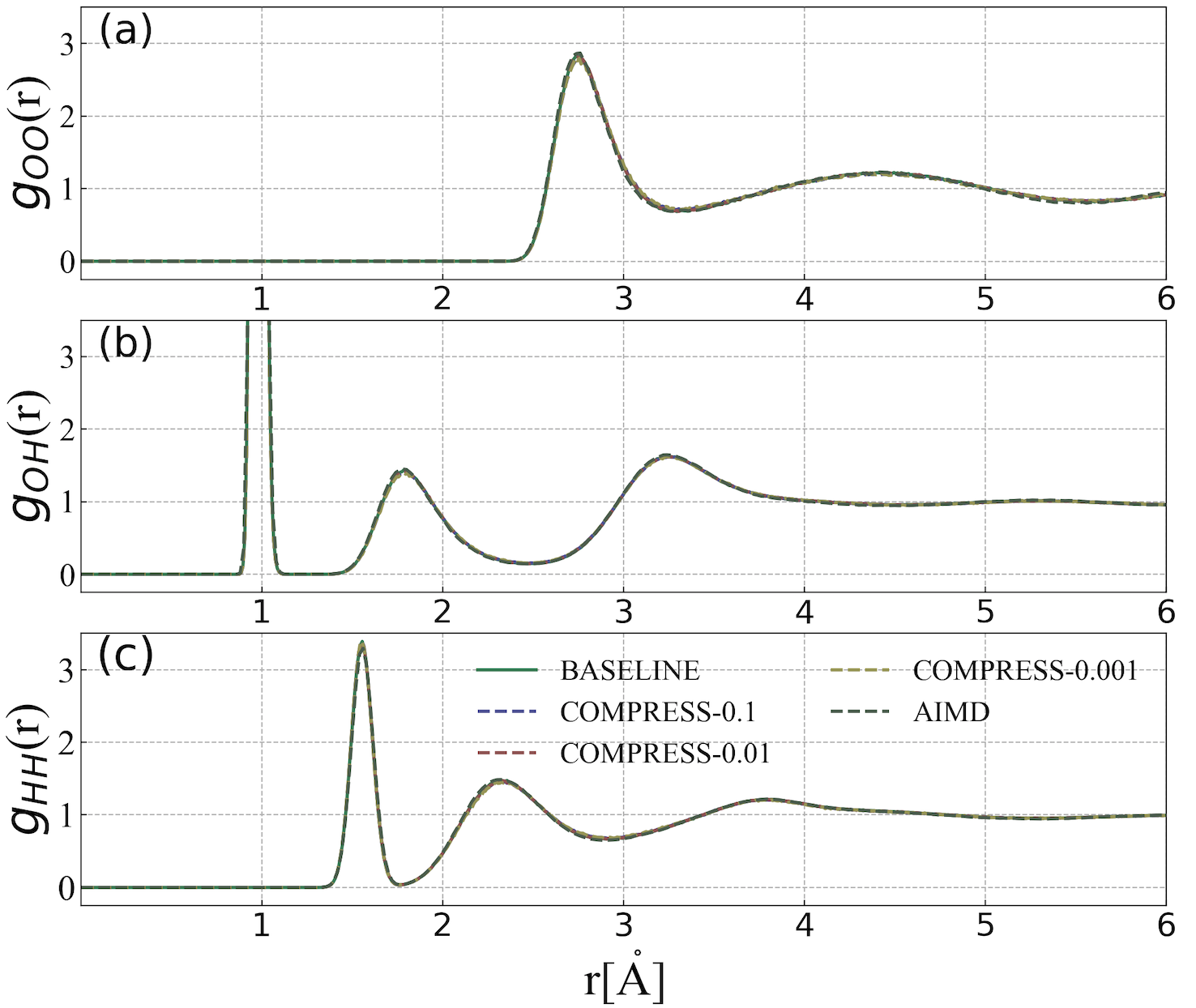}
 \caption{\label{fig:water-rdf}
 Radial distribution functions $g_{OO}(r)$, $g_{OH}(r)$, and
$g_{HH}(r)$ of liquid water at ambient conditions. The methods used include the {\it ab initio} MD~\cite{distasio2014jcp} and four DP models: the original DP model (DP BASELINE) and the compressed DP model (DP COMPRESS) with tabulation steps being 0.1, 0.01, and 0.001.
The DP data are collected from 100-ps MD simulations of 512 water molecules with a time step of 0.5 fs.}
\end{figure}

Fig.~\ref{fig:rmse-properties} demonstrates the general effect of the tabulation step on these properties under both single and double precisions.
Detailed results demonstrating each data point and the error distribution can be found in Fig. S1.
While the results under single precision are less satisfactory due to the limitation of significant digits, decreasing the tabulation step under the double precision reasonably improves the accuracy.
The results suggest that $\Delta t=$0.01 or smaller tabulation steps under double precision could ensure reliable results for the tested properties.
Negligible differences in most cases are observed in such occasions, and in particular, the root-mean-square error (RMSE) of formation energies compared with the original DP model is smaller than the accuracy limit of double precision.

We find that the RMSE of elastic moduli is higher than other properties by using the double precision.
This is caused by the presence of one and two outlier points when $\Delta t=$0.001 and $\Delta t=$0.01 are adopted in DP Compress, respectively. Despite the overall excellent agreement, which yields absolute errors smaller than $10^{-3}$ GPa for all of the rest data points.
The outliter is caused by the high instability of the initial deformed structure used to calculate the elastic property with a finite-difference method.
Therefore, this feature implies that detailed tests should be performed before the usage of DP Compress for extensive studies.
We suggest that the tests should cover properties related to the high-order derivatives.

Besides the above solid properties, we further conduct MD simulations for liquid water in order to validate the DP compress scheme for liquid systems. The system contains 512 water molecules. We ran 100-ps DeePMD simulations with a time step of 0.5 fs.
We compute the structural property of liquid water by carrying out MD simulations.
As shown in Fig.~\ref{fig:water-rdf}, the O-O, O-H, and H-H radial distribution functions predicted by all of the compressed models with different tabulation steps agree fairly well with the baseline results. In particular, the resulting radial distribution functions also agree well with the baseline data even when the tabulation step is set to as large as 0.1. This implies that the structures of liquid water can be well addressed by using different lengths of the tabulation step in the DP compress scheme.

\begin{figure}
 \centering
 \includegraphics[width=0.96\textwidth]{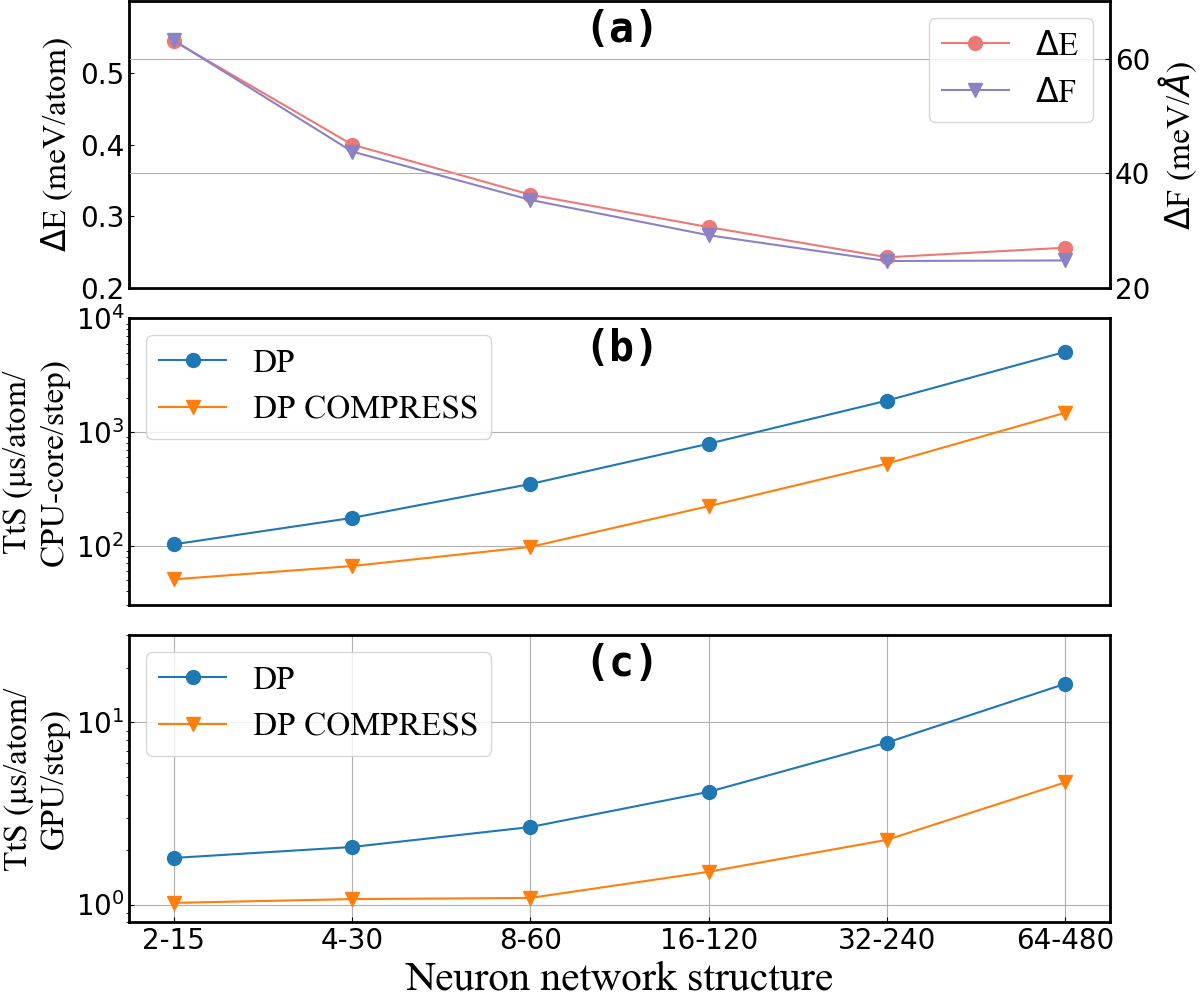}
 \caption{
 \label{fig:nn-performance}
 Accuracy and time-to-solution (TtS) of DP models by using CPU and GPU devices with different neuron network structures.
 (a) Root mean square errors of energy and forces as a function of different neuron network structures. (b) Time-to-solution ($\upmu$s/atom/CPU-core/step) of DP models tested on a 6-core Intel Xeon 8163 CPU.
 (c) Time-to-solution ($\upmu$s/atom/GPU/step) of DP models tested on an NVIDIA V100 GPU.
 The labels on the $x$-axis, $a$-$b$ depict the embedding net with layer sizes of $a$-2$a$-4$a$, and the fitting net with layer sizes of $b$-$b$-$b$.
 }
\end{figure}

The performances of the original and the compressed models depend on the detailed structures of the embedding net and the fitting net. To clarify this, we test a system with 4096 water molecules and the results are shown in Fig.~\ref{fig:nn-performance}.
The embedding net structure is chosen to be $a$-$2a$-$4a$ with $a$ being the number of neurons, while the fitting net structure is set to $b$-$b$-$b$ with $b$ being the number of neurons. Generally speaking, a larger number of $a$ or $b$ leads to more accurate models but a lower efficiency.
Notably, as illustrated in Fig.~\ref{fig:nn-performance},
we observe that the accuracy of DP models saturates when $a=32$ and $b=240$.
In this regard, we suggest that the best network structure may be case-specific and depends on the quality and complexity of the training dataset.
Furthermore, the conventional compression schemes adopted in the machine learning community, such as pruning and neural architecture search, should benefit this process, and will be considered in the future.

With the abovementioned techniques implemented in the compressed model, the computational hotspot of the compressed DP model has changed, which has profound implications. On one hand, since the embedding net is the major computational bottleneck in the original DP model, a better efficiency can be gained only if the size of the embedding net is reduced. On the other hand,
in the compressed DP model, the computational costs for generating the descriptor and for using the fitting net are comparable.
Therefore, reducing the number of outputs ($M$) in the embedding net or reduce the size of the fitting net can lead to a better efficiency.

\section{Conclusion}
In summary, we propose a DP Compress scheme that can significantly boost the performance of DP models with controllable loss of accuracy. 
\MC{We suggest that using double precision and a tabulation step of 0.01 or smaller for the DP Compress scheme.}
The new scheme will benefit all of the users of DeePMD-kit, as well as inspire other methodology developers in the field of machine learning assisted scientific computing.
In the future, more optimizations on different operators, on the computational graph, and on multiple hardware devices, would be needed.
Moreover, DP Compress can be generalized to other DP-based models, such as vectors~\cite{zhang2020dw} and tensors~\cite{grace2020raman}, without essential difficulties.
With DeePMD-kit being an open-source software package, we expect that more innovative and useful schemes can be continuously integrated in the code by developers not limited to the authors, and we expect that all these later improvements can benefit the users in a timely manner.

\MC{
We suggest that the best hyper-parameters for a compressed model are case-specific, which typically depend on the complexity and quality of the training data. 
For most applications, the default settings for these parameters in the DeePMD-kit package can be used without tuning.
However, if one wants to achieve a better accuracy under a certain computational budget, 
he/she has to carefully tune the accuracy- and efficiency-relevant  hyper-parameters.
Here we provide some guidances for tuning the relevant hyper-parameters.
(1) A larger cutoff radius usually leads to a better accuracy, but the model may become more difficult to train. Note that both training and inference costs increase with the cutoff radius.
(2) Larger sizes of the embedding and fitting networks imply a better accuracy but higher computational costs are expected.
(3) A larger number of training steps often results in generating a model with better accuracy. Although the training cost growth in proportional to the number of training steps, the inference operations and MD costs are invariant.
(4) A smaller tabulation step size leads to a higher accuracy. 
The memory costs grow in proportion to the inverse of the step size, while the computational costs only marginally increase: 
reducing the tabulation step size by 10 times leads to an increment of less than 5\% in the time-to-solution. 
}

$\\$
{\bf Acknowledgements}
$\\$
The work of D. Lu and M. Chen is supported by the National Science Foundation of China under Grant No.12122401 and 12074007.
Y. Chen is supported by the DOE Award DE-SC0019394 (Center Chemistry in Solution and at Interfaces).
L. Zhang is supported by Beijing Academy of Artificial Intelligence(BAAI).
W. Jia is supported by Institute of Computing Technology under Grant No. CARCH5101 and 55E061100.
The work of H.W.~was supported by the National Science Foundation of China under Grant No.11871110 and 12122103.
Part of the numerical simulations were performed on the High Performance Computing Platform of CAPT.

$\\$
\MC{{\bf Data and Code Availability}}
$\\$
\MC{The training data sets and the DP models for the Cu, H$_2$O, and Al-Cu-Mg systems can be found on the DP-Library 
(Cu: \url{https://dplibrary.deepmd.net/\#/project\_details?project\_id=202010.008}, H$_2$O: \url{https://dplibrary.deepmd.net/\#/project\_details?project\_id=202010.001} and Al-Cu-Mg: \url{https://dplibrary.deepmd.net/\#/project\_details?project\_id=202010.002}).
The DP Compress code is publicly available at \url{https://github.com/deepmodeling/deepmd-kit}.
}

$\\$
{\bf Author contributions}
$\\$
D.L. implemented the method. D.L., W.J. and Y.C. carried out the simulations and performed the analysis.
L.Z., W.J., H.W. and M.C. designed the project.
All authors contributed to the discussions and revisions of the manuscript.

$\\$
{\bf Conflict of Interest}
$\\$
The authors declare no competing interests.

\newpage

\bibliography{ref}{}
\bibliographystyle{unsrt}

\end{document}


\titleformat{\section}{\centering\bfseries\large}{\thesection}{1em}{}
\section*{SUPPLEMENTARY MATERIALS}
\vspace{10mm}
\title{\MC{DP Compress: a Model Compression Scheme for Generating Efficient Deep Potential Models}
  }

\author{Denghui Lu}
\affiliation{HEDPS, CAPT, College of Engineering, Peking University, Beijing 100871, P.R.~China}

\author{Wanrun Jiang}
\affiliation{Songshan Lake Materials Laboratory, Dongguan, Guangdong 523808, P.R.~China}
\affiliation{Institute of Physics, Chinese Academy of Sciences, Beijing 100190, P.R.~China}

\author{Yixiao Chen}
\affiliation{Program in Applied and Computational Mathematics, Princeton University, Princeton, NJ, USA}

\author{Linfeng Zhang}
\affiliation{Beijing Institute of Big Data Research, Beijing 100871, P.R.~China}

\author{Weile Jia}
\affiliation{Institute of Computing Technology, Chinese Academy of Sciences, Beijing 100190, P.R.~China}
\affiliation{University of Chinese Academy of Sciences, Beijing 100049, P.R.~China}

\author{Han Wang}
\affiliation{Laboratory of Computational Physics,
  Institute of Applied Physics and Computational Mathematics, Fenghao East Road 2, Beijing 100094, P.R.~China}
\affiliation{HEDPS, CAPT, College of Engineering, Peking University, Beijing 100871, P.R.~China}

\author{Mohan Chen}
\email{mohanchen@pku.edu.cn}
\affiliation{HEDPS, CAPT, College of Engineering, Peking University, Beijing 100871, P.R.~China}

\maketitle
\titleformat{\section}{\flushleft\bfseries\large}{\thesection}{1em}{}
\newpage
\setlength{\parskip}{.5em}
\vspace{1mm}
\renewcommand\thefigure{S\arabic{figure}}
\setcounter{figure}{0}
\let\cleardoublepage\clearpage
\section{\MC{Results from Cu and Al-Cu-Mg Models}}

\begin{figure}[H]
    \centering   
        \subfloat
    {
        \begin{minipage}[t]{0.23\textwidth}
            \centering
            \includegraphics[width=1.0\textwidth]{figs/forme_double_0.1.eps} 
        \end{minipage}
        \begin{minipage}[t]{0.23\textwidth}
            \centering
            \includegraphics[width=1.0\textwidth]{figs/vol_doubel_0.1.eps} 
        \end{minipage}
        \begin{minipage}[t]{0.23\textwidth}
            \centering
            \includegraphics[width=1.0\textwidth]{figs/elastic_double_0.1.eps} 
        \end{minipage}
        \begin{minipage}[t]{0.23\textwidth}
            \centering
            \includegraphics[width=1.0\textwidth]{figs/surf_double_0.1.eps} 
        \end{minipage}
    }

        \subfloat
    {
        \begin{minipage}[t]{0.23\textwidth}
            \centering
            \includegraphics[width=1.0\textwidth]{figs/forme_double_0.01.eps} 
        \end{minipage}
        \begin{minipage}[t]{0.23\textwidth}
            \centering
            \includegraphics[width=1.0\textwidth]{figs/vol_doubel_0.01.eps} 
        \end{minipage}
        \begin{minipage}[t]{0.23\textwidth}
            \centering
            \includegraphics[width=1.0\textwidth]{figs/elastic_double_0.01.eps} 
        \end{minipage}
        \begin{minipage}[t]{0.23\textwidth}
            \centering
            \includegraphics[width=1.0\textwidth]{figs/surf_double_0.01.eps} 
        \end{minipage}
    }
     
             \subfloat
    {
        \begin{minipage}[t]{0.23\textwidth}
            \centering
            \includegraphics[width=1.0\textwidth]{figs/forme_double_0.001.eps} 
        \end{minipage}
        \begin{minipage}[t]{0.23\textwidth}
            \centering
            \includegraphics[width=1.0\textwidth]{figs/vol_doubel_0.001.eps} 
        \end{minipage}
        \begin{minipage}[t]{0.23\textwidth}
            \centering
            \includegraphics[width=1.0\textwidth]{figs/elastic_double_0.001.eps}
        \end{minipage}
        \begin{minipage}[t]{0.23\textwidth}
            \centering
            \includegraphics[width=1.0\textwidth]{figs/surf_double_0.001.eps} 
        \end{minipage}
    }
    
            \subfloat
    {
        \begin{minipage}[t]{0.23\textwidth}
            \centering
            \includegraphics[width=1.0\textwidth]{figs/forme_single_0.1.eps} 
        \end{minipage}
        \begin{minipage}[t]{0.23\textwidth}
            \centering
            \includegraphics[width=1.0\textwidth]{figs/vol_single_0.1.eps} 
        \end{minipage}
        \begin{minipage}[t]{0.23\textwidth}
            \centering
            \includegraphics[width=1.0\textwidth]{figs/elastic_single_0.1.eps} 
        \end{minipage}
        \begin{minipage}[t]{0.23\textwidth}
            \centering
            \includegraphics[width=1.0\textwidth]{figs/surf_single_0.1.eps} 
        \end{minipage}
    }
     
            \subfloat
    {
        \begin{minipage}[t]{0.23\textwidth}
            \centering
            \includegraphics[width=1.0\textwidth]{figs/forme_single_0.01.eps} 
        \end{minipage}
        \begin{minipage}[t]{0.23\textwidth}
            \centering
            \includegraphics[width=1.0\textwidth]{figs/vol_single_0.01.eps} 
        \end{minipage}
        \begin{minipage}[t]{0.23\textwidth}
            \centering
            \includegraphics[width=1.0\textwidth]{figs/elastic_single_0.01.eps} 
        \end{minipage}
        \begin{minipage}[t]{0.23\textwidth}
            \centering
            \includegraphics[width=1.0\textwidth]{figs/surf_single_0.01.eps} 
        \end{minipage}
    }   
    
            \subfloat
    {
        \begin{minipage}[t]{0.23\textwidth}
            \centering
            \includegraphics[width=1.0\textwidth]{figs/forme_single_0.001.eps} 
        \end{minipage}
        \begin{minipage}[t]{0.23\textwidth}
            \centering
            \includegraphics[width=1.0\textwidth]{figs/vol_single_0.001.eps} 
        \end{minipage}
        \begin{minipage}[t]{0.23\textwidth}
            \centering
            \includegraphics[width=1.0\textwidth]{figs/elastic_single_0.001.eps} 
        \end{minipage}
        \begin{minipage}[t]{0.23\textwidth}
            \centering
            \includegraphics[width=1.0\textwidth]{figs/surf_single_0.001.eps} 
        \end{minipage}
    }
    \caption{\label{fig:detailed_properties}Comparison of formation energy, equilibrium volume, elastic moduli, and unrelaxed surface formation energy predicted by the original DP model and the compressed DP model, using different precision setups (single and double), as well as different tabulation steps (0.1, 0.01, 0.001). We use 64 structures, including 6 structures of Cu and 58 binary and ternary alloy structures from the Materials Projects (MP) database. Dased lines in figures for formation energy, equilibrium volume, elastic modul and unrelaxed surface formation energy are respectively $y = x \pm 0.03$, $y = x \pm 0.4$, $y = x \pm 10$ and $y = x \pm 0.1$. The errors smaller than $10^{-16}$ in the inserted figures of (a) and (d) are truncated by considering the limit of significant digits under the double precision.}
    
\end{figure}

Detailed comparison of predictions from DP Compress with those from the original model are demonstrated in Fig.~\ref{fig:detailed_properties}.
Three tabulation steps (0.1, 0.01 and 0.001) with two numerical precision (double and single) are listed.
By utilizing the double precision, improvement is shown when switching from  $\Delta t=$0.1 (double-0.1) to $\Delta t=$0.01 (double-0.01), where the range for the error distribution of the formation energy even decreases from the order of $\rm 10^{-3} ~ eV{\cdot}atom ^{-1}$ to beneath $\rm 10^{-16} ~ eV{\cdot}atom ^{-1}$.
Both $\Delta t=$0.01 and 0.001 (double-0.001) present satisfactory results in most cases.

However, two oulier points with errors around -14.8 (4.6) GPa in the bulk (shear) modulus are shown for the elastic moduli from the double-0.01 DP Compress data.
Though all of the rest 126 data points yield absolute errors smaller than $10^{-3}$ GPa, contributions from the two outlier points result in the relatively high RMSE in Fig. 4(a).
Similarly, an outlier point with the error about 14.1 Gpa is shown for the elastic moduli from double-0.001.
Consequently, the RMSE for the elastic moduli under double precision in Fig. 4(a) seems to be less improved, when decreasing the tabulation step.
In conclusion, both double-0.01 and double-0.001 parameters yield remarkably better elastic results than double-0.1 as shown in Fig.~\ref{fig:detailed_properties}.

\section{\MC{Computational Setups for the DP models}}
\MC{We list the computational setups for generating the Cu, H$_2$O, and Al-Cu-Mg DP potentials, which were adopted from previous works~\cite{zhang2020dpgen,zhang2021phase,jiang2021accurate,Jia2020GB}. The Adam stochastic gradient descend method was used for all of the models. In addition, We describe how we split the training, validation, and test data sets for the above three systems. Note that all of the training data were obtained via the concurrent learning platform named DP-GEN~\cite{zhang2020dpgen}, no data were used for validation during the training process.}

\MC{For the Cu DP model~\cite{zhang2020dpgen,Jia2020GB}, the configuration space was chosen with certain conditions, i.e., the temperature ranges from 0 K to 2 $T_m$ and the pressure ranges from 1 to 50000 bar, where $T_m$ = 1357.77 K denotes the melting point of Cu at the ambient conditions.
DPMD simulations started form bulk and surface initial configurations, which were generated from the randomly perturbated face-centered cubic (fcc), hexagonal-closed-packed (hcp) and body-centered cubic (bcc) crystal structures.
%
The learning rate was chosen to be $10^{-3}$ and exponentially decayed to 3.5$\times 10^{-8}$ after $4 \times 10^5$ training steps. Once the DP-GEN iterations were converged, production models were trained with the training steps being 8$\times$10$^6$.}

\MC{In terms of the H$_2$O DP model~\cite{zhang2021phase}, 36 concurrent iterations were adopted to train the model with temperatures ranging from low temperature to about 2400 K and pressure up to 50 GPa. The accumulated number of snapshots in the training dataset was 31058. The number of training steps was set to $10^6$. The starting learning rate was set to 0.002 and exponentially decayed to 7.4$\times$10$^{-8}$ at the end of the training.}

\MC{Regarding the Al-Cu-Mg DP model, the whole concentration space was explored, i.e., Al$_x$Cu$_y$Mg$_z$ with $0\leq x,y,z\leq 1$, $x+y+z=1$. The configuration space was chosen with temperatures ranging from around 50.0 to 2579.8 K and the pressure ranging from 1 to 50000 bar. 
The initial configurations were generated from the fcc, hcp and bcc structures.
%
The learning rate was chosen to be $10^{-3}$ and exponentially decayed to 3.5$\times 10^{-8}$ after $10^6$ training steps. Production models were trained with the training steps being 1.6$\times$10$^7$ once the DP-GEN iterations were converged.}


\MC{The test datasets were individually constructed, for the Cu and Al-Cu-Mg systems, a variety of crystal structures from the Materials Project were selected, which were not included in the training process. To be specific, we adopted 6 crystal structures of Cu and 58 crystal structures of binary and ternary alloys including Al, Mg, and Cu elements from the Materials Project Database. In detail, the 
Materials Project ID numbers used in this work are as follows: mp-30, mp-1010136, mp-1059259, mp-989695
mp-989782, mp-998890 for Cu; mp-1008555, mp-1022721, mp-10886, mp-1182885, mp-1183161, mp-1183216, mp-1228752, mp-12777, mp-12794, mp-12802, mp-2500, mp-593, mp-985806, mp-985825, mp-998 for Al-Cu; mp-1016271, mp-1038779, mp-1038818, mp-1038916, mp-1038934, mp-1039010, mp-1039019, mp-1039119, mp-1039141, mp-1039180, mp-1039192, mp-1094116, mp-1094664, mp-1094666, mp-1094683, mp-1094685, mp-1094692, mp-1094700, mp-1094961, mp-1094970, mp-1094987, mp-1102064, mp-1185596, mp-1194114, mp-1222013, mp-12766, mp-17659, mp-2151, mp-568106, mp-978271 for Al-Mg;
mp-1038, mp-1185803, mp-1185856, mp-2481, mp-865014, mp-972916, mp-978279 for Cu-Mg; and mp-1200279, mp-1222119, mp-1222208, mp-1222214, mp-30178, mp-3034 for Al-Cu-Mg.
In order to test the H$_2$O DP model ~\cite{zhang2021phase}, 5141 configurations were selected from 67 isothermal-isobaric DPMD trajectories from relevant thermodynamic domain.
These configurations form an independent dataset to yield the error of the DP model relative to density functional theory calculations.
}

\newpage

\bibliography{ref}{}
\bibliographystyle{unsrt}